\documentclass[twocolumn]{aastex631}
\received{x}
\revised{x}
\accepted{x}
\submitjournal{ApJ}
\usepackage{fancyvrb} 
\usepackage{breakurl}
\usepackage{multirow}
\usepackage{xcolor}
\VerbatimFootnotes    

\begin{document}

\newcommand{\vdag}{(v)^\dagger}
\newcommand\aastex{AAS\TeX}
\newcommand\latex{La\TeX}

\shorttitle{Multimessenger Picture of J1048+7143}
\shortauthors{Kun et al.}
\graphicspath{{./}{figures/}}

\title{Multimessenger Picture of J1048+7143\footnote{Released xxxx 2022}}

\correspondingauthor{Emma Kun}
\email{kun.emma@csfk.org}

\author[00000-0003-2769-3591]{Emma Kun}
\affiliation{Konkoly Observatory, ELKH Research Centre for Astronomy and Earth Sciences, Konkoly Thege Miklós \'ut 15-17, H-1121 Budapest, Hungary}
\affiliation{CSFK, MTA Centre of Excellence, Konkoly Thege Miklós \'ut 15-17, H-1121 Budapest, Hungary}

\author[00000-0000-0000-0000]{Ilja Jaroschewski}
\affiliation{Theoretical Physics IV: Plasma-Astroparticle Physics, Faculty for Physics \& Astronomy, Ruhr University Bochum, 44780 Bochum, Germany}
\affiliation{Ruhr Astroparticle And Plasma Physics Center (RAPP Center), Ruhr-Universit\"at Bochum, 44780 Bochum, Germany}

\author[00000-0000-0000-0000]{Armin Ghorbanietemad}
\affiliation{Theoretical Physics IV: Plasma-Astroparticle Physics, Faculty for Physics \& Astronomy, Ruhr University Bochum, 44780 Bochum, Germany}
\affiliation{Ruhr Astroparticle And Plasma Physics Center (RAPP Center), Ruhr-Universit\"at Bochum, 44780 Bochum, Germany}

\author[0000-0003-3079-1889]{S\'andor Frey}
\affiliation{Konkoly Observatory, ELKH Research Centre for Astronomy and Earth Sciences, Konkoly Thege Miklós \'ut 15-17, H-1121 Budapest, Hungary}
\affiliation{CSFK, MTA Centre of Excellence, Konkoly Thege Miklós \'ut 15-17, H-1121 Budapest, Hungary}
\affiliation{Institute of Physics, ELTE E\"{o}tv\"{o}s Lor\'{a}nd University, P\'{a}zm\'{a}ny P\'{e}ter s\'{e}t\'{a}ny 1/A, H-1117 Budapest, Hungary}

\author[0000-0002-1748-7367]{Julia Becker Tjus}
\affiliation{Theoretical Physics IV: Plasma-Astroparticle Physics, Faculty for Physics \& Astronomy, Ruhr University Bochum, 44780 Bochum, Germany}
\affiliation{Ruhr Astroparticle And Plasma Physics Center (RAPP Center), Ruhr-Universit\"at Bochum, 44780 Bochum, Germany}

\author[00000-0000-0000-0000]{Silke Britzen}
\affiliation{Max-Panck-Institut für Radioastronomie, Auf dem Hügel 69, 53121 Bonn, Germany}

\author[0000-0003-1020-1597]{Krisztina \'Eva Gab\'anyi}
\affiliation{Department of Astronomy, ELTE E\"{o}tv\"{o}s Lor\'{a}nd University, P\'{a}zm\'{a}ny P\'{e}ter s\'{e}t\'{a}ny 1/A, H-1117 Budapest, Hungary}
\affiliation{ELKH-ELTE Extragalactic Astrophysics Research Group, P\'{a}zm\'{a}ny P\'{e}ter s\'{e}t\'{a}ny 1/A, H-1117 Budapest, Hungary}
\affiliation{Konkoly Observatory, ELKH Research Centre for Astronomy and Earth Sciences, Konkoly Thege Miklós \'ut 15-17, H-1121 Budapest, Hungary}
\affiliation{CSFK, MTA Centre of Excellence, Konkoly Thege Miklós \'ut 15-17, H-1121 Budapest, Hungary}

\author[0000-0003-1254-0197]{Vladimir Kiselev}
\affiliation{Theoretical Physics IV: Plasma-Astroparticle Physics, Faculty for Physics \& Astronomy, Ruhr University Bochum, 44780 Bochum, Germany}
\affiliation{Ruhr Astroparticle And Plasma Physics Center (RAPP Center), Ruhr-Universit\"at Bochum, 44780 Bochum, Germany}

\author[00000-0000-0000-0000]{Leander Schlegel}
\affiliation{Theoretical Physics IV: Plasma-Astroparticle Physics, Faculty for Physics \& Astronomy, Ruhr University Bochum, 44780 Bochum, Germany}
\affiliation{Ruhr Astroparticle And Plasma Physics Center (RAPP Center), Ruhr-Universit\"at Bochum, 44780 Bochum, Germany}

\author[00000-0000-0000-0000]{Marcel Schroller}
\affiliation{Theoretical Physics IV: Plasma-Astroparticle Physics, Faculty for Physics \& Astronomy, Ruhr University Bochum, 44780 Bochum, Germany}
\affiliation{Ruhr Astroparticle And Plasma Physics Center (RAPP Center), Ruhr-Universit\"at Bochum, 44780 Bochum, Germany}

\author[0000-0003-4513-8241]{Patrick Reichherzer}
\affiliation{Theoretical Physics IV: Plasma-Astroparticle Physics, Faculty for Physics \& Astronomy, Ruhr University Bochum, 44780 Bochum, Germany}
\affiliation{Ruhr Astroparticle And Plasma Physics Center (RAPP Center), Ruhr-Universit\"at Bochum, 44780 Bochum, Germany}
\affiliation{IRFU, CEA, Université Paris-Saclay, F-91191 Gif-sur-Yvette, France}

\author[0000-0003-0721-5509]{Lang Cui}
\affiliation{Xinjiang Astronomical Observatory, Chinese Academy of Sciences, 150 Science 1-Street, Urumqi 830011, China}

\author{Xin Wang}
\affiliation{Xinjiang Astronomical Observatory, Chinese Academy of Sciences, 150 Science 1-Street, Urumqi 830011, China}

\author{Yuling Shen}
\affiliation{Xinjiang Astronomical Observatory, Chinese Academy of Sciences, 150 Science 1-Street, Urumqi 830011, China}



\begin{abstract}
We draw the multimessenger picture of J1048+7143, a flat-spectrum radio quasar known to show quasi-periodic oscillations in the $\gamma$-ray regime. We generate the adaptively-binned Fermi Large Area Telescope light curve of this source above 168 MeV to find three major $\gamma$-ray flares of the source, such that all three flares consist of two-two sharp sub-flares. Based on radio interferometric imaging data taken with the Very Large Array, we find that the kpc-scale jet is directed towards west, while our analysis of $8.6$-GHz very long baseline interferometry data, mostly taken with the Very Long Baseline Array, revealed signatures of two pc-scale jets, one pointing towards east, one pointing towards south. We suggest that the misalignment of the kpc- and pc-scale jets is a revealing signature of jet precession. We also analyze the $5$-GHz total flux density curve of J1048+7143 taken with the Nanshan(Ur) and RATAN-600 single dish radio telescopes and find two complete radio flares, slightly lagging behind the $\gamma$-ray flares. We model the timing of $\gamma$-ray flares as signature of the spin--orbit precession in a supermassive black hole binary, and find that the binary could merge in the next $\sim 60-80$ years. We show that both the Pulsar Timing Arrays and the planned Laser Interferometer Space Antenna lack sensitivity and frequency coverage to detect the hypothetical supermassive black hole binary in J1048$+$7143. We argue that the identification of sources similar to J1048+7143 plays a key role to reveal periodic high-energy sources in the distant Universe.
\end{abstract}


\section{Introduction}
\label{section:intro}

Active galactic nuclei (AGN) are among the most luminous persistent objects in the Universe. The activity of the galactic nucleus is driven by material captured by a supermassive black hole (SMBH) with up to billions of Solar masses. Part of this material is accelerated to speeds close to the speed of light and expelled from the immediate vicinity of the black hole in the form of a pair of relativistic particle jets \citep[e.g.][]{Begelman1984}. The Blandford--Znajek \citep[][]{BlandfordZnajek1977} mechanism allows to extract rotational energy from a black hole due to the dragging of magnetic field lines, and such, jets can feed upon the massive energy reservoir of rotating black holes. In fact, AGN jets are the most powerful particle accelerators in the Universe.

The very long baseline interferometry (VLBI) technique at radio frequencies allows us to observe radio-emitting celestial objects with angular resolution of milliarcsec (mas) or below. With such high resolution, the jet evolution of radio-loud AGN can be studied typically on pc-scale, even over decade-long time-scales if monitoring data are available. AGN jets show a variety of motions. The jet blobs in some of them move outward from the core \citep[e.g.][and references therein]{Kun2014}, in some of them they are oscillating \citep[e.g.][]{Britzen2010,Kun2018}. Some components show linear motion, and others show acceleration. Components can move radially and non-radially, too \citep[e.g.][]{Lister2019}. What we see is likely the composition of inner jet features combined with geometric effects \citep[e.g.][]{Britzen2010}. 

According to the hierarchical galaxy evolution models \citep[e.g.,][]{Cole2000}, galaxies gather their masses through major merger phases. As a result, million to billion solar mass SMBHs formed at the center of massive galaxies through a sequence of accretion and merger phases \citep[e.g.][]{Malbon2007,Capelo2015}. Some of them might be members of binary systems due to the long time-scale of the merger \citep[e.g.][]{SearleZinn1978,Begelman1980,Komossa2006}. When a member of the supermassive binary black hole (SMBBH) is radio-loud due to its relativistic jets -- approximately $10\%$ of single AGN qualify as jetted \citep[e.g.][]{Ivezic2002} --, a periodic jet structure might indicate that the jet emitter black hole interacts gravitationally with another black hole \citep[e.g.][]{Roos1993,Britzen2001,Valtonen2012,Kun2014,Kun2015}. In the inspiral phase of the merger \citep[e.g.][]{MerrittMilos2005}, periodic structures might arise due to the spin--orbit precession of the jet emitting black hole \citep{GerPLB2009}. The post-Newtonian dynamics of the close black hole binaries is introduced in detail in \citet{Barker1975,Barker1979,Kidder1995,Gergely2010a,Gergely2010b}. Such massive binary black holes have sub-pc separations, and their identification is possible only through implicit methods, like the modeling of periodic jets. The future space-based gravitational wave (GW) observatory, Laser Interferometer Space Antenna \citep[LISA,][]{LISA2017_WP} will operate in the low-frequency range, between 0.1 mHz and 1 Hz, where it will be able to capture the gravitational wave signal of inspiralling and merging massive black holes (with $10^3 - 10^7 \mathrm{M}_\odot$).

In some cases, precessing jets might be able to produce high-energy neutrino emissions, when the moving jet interacts with a strong radiation field and/or dense region of matter. These can be either the external photon field of the accretion disk \citep[e.g.][]{Hoerbe2020}, or a second jet in a binary system \citep[e.g.][]{Britzen2019}, the broad-line region of the AGN \citep[e.g.][]{Britzen2021} etc. High-energy neutrinos would serve as unambiguous proof of particle acceleration from such sources, in contrast to the direct messengers, as charged cosmic rays are deflected on their way and do not reveal individual sources. Gamma-ray signatures are ambiguous, as leptonic processes can contribute as well. Neutrinos, on the other hand, have the disadvantage of being difficult to detect. In combination, looking at the multimessenger picture of active galaxies is the only way to understand the high-energy emission from these sources \citep{Becker2008}. AGN can even be observed from enormous distances corresponding to only a few percent of the current age of the Universe \citep[e.g.][]{Frey2010,Gabanyi2015,Perger2017,Krezinger2022}. High-energy $\gamma$-rays from such distances are attenuated on their way to Earth \citep[e.g.][]{Dermer2009,Gilmore2012}, cosmic rays are scrambled by cosmic magnetic fields \citep[e.g.][]{Reichherzer2022}, and therefore observation of (quasi-)periodic neutrino emission might be the only direct evidence of periodic high-energy particle accelerators in the sky. Therefore it is important to find such sources at moderate redshifts, from where we still can observe the periodic flaring via high-energy photons. Joint observation of this behavior with signs of jet precession at radio wavelengths might reveal high-energy neutrino factories of massive binary black holes.

Assuming the three years elapsed between the detection of the 2014/2015 large neutrino flare \citep{ICTXS2018b} and the detection of IC170922A \citep{ICTXS2018a} by the Antarctic IceCube Neutrino Observatory, \citet{deBruijn2020} showed that if the identified neutrino source TXS~0506+056 harbors an SMBBH, then the next neutrino flare could already have occurred, possibly still hidden in IceCube’s not-yet-analyzed data. They derived the binary properties that would lead to the detection of gravitational waves from this system by LISA over the next decade. Their results connected for the first time the possible neutrino and gravitational-wave signatures of such sources. 

Quasi-periodic oscillations (QPOs) in many blazars can be observed in optical, X-ray, radio bands, as well as sometimes in the $\gamma$-ray regime \citep[e.g.][]{Ren2022}. Recently, \citet{Wang2022} obtained 5-day binned light curves for J1048+7143 (S5~1044+71) with a time coverage of $\sim12.9$~yr, based on Fermi Large Area Telescope (LAT) data. The LAT instrument onboard the Fermi Gamma-ray Space Telescope is designed to cover the energy band from 20 MeV to above 300 GeV. Applying five different methods, \citet{Wang2022} found a possible QPO with a period of $3.06 \pm 0.43$~yr at the significance level of $\sim3.6\sigma$. They concluded that the quasi-periodic flaring activity suggests the presence of a SMBBH in J1048+7143.

\begin{figure*}
    \centering
    \includegraphics[angle=270,scale=0.65]{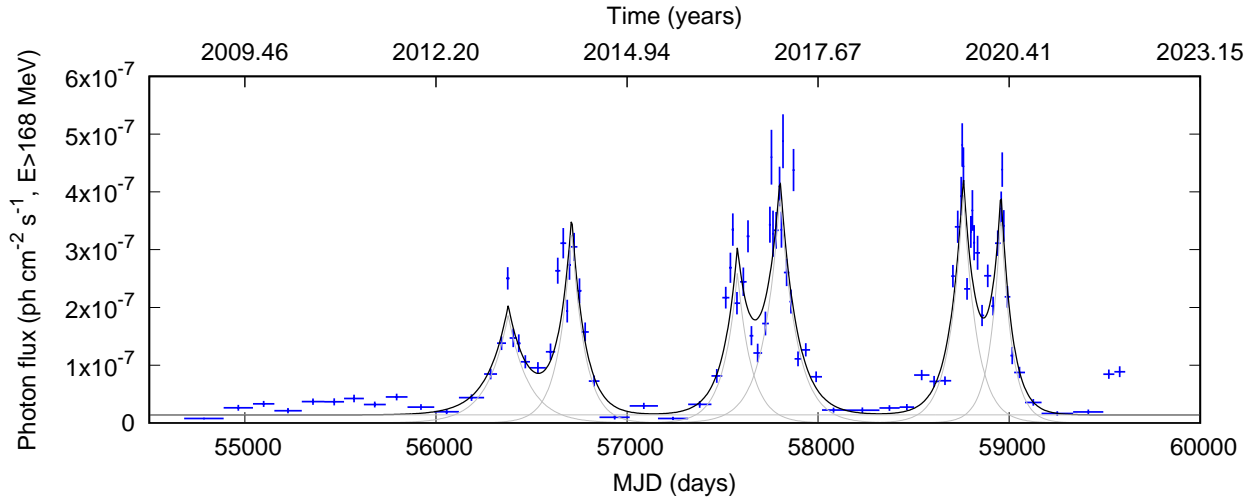}
    \caption{The Fermi-LAT photon flux ($168\,\mathrm{MeV} < E < 800\,\mathrm{GeV}$) of 4FGL~J1048.4+7143 as a function of time (black points with error bars). }
    \label{fig:gammalc}
\end{figure*}

J1048+7143 is a low-synchrotron-peaked flat-spectrum radio quasar (FSRQ) at redshift $z=1.15$ \citep{Polatidis1995}, associated with the Fermi $\gamma$-ray source 4FGL~1048.4+7143 \citep{Fermi4fgldr12020,Fermi4fgldr22020}. In this paper, based on our analysis of Fermi-LAT $\gamma$-ray and VLBI radio observations, we propose the multimessenger picture of J1048+7143 by assuming that the flaring activity in these regimes is due to the spin--orbit precession of a binary black hole at the heart of this FSRQ. We expand the precession model of \citet{deBruijn2020}, and predict the detection time of the next $\gamma$-ray flare from J1048+7143, assuming that the quasi-periodicity of its $\gamma$-ray light curve is due to the spin--orbit precession at the jet base.

In Section~\ref{sec:gamma}, we present our analysis of Fermi-LAT $\gamma$-ray observations, generate the light curve using a likelihood fitting technique to obtain the individual flux data points, and model the spectral energy distribution (SED) of 4FGL~J1048.4+7143 (J1048+7143). In Section~\ref{sec:radio}, we analyze archival VLBI imaging observations of J1048+7143 (mostly VLA and VLBA) to study the radio jet structure on kpc and pc scales. We also analyze single-dish $5$-GHz band total flux density monitoring observations by the Nanshan and RATAN-600 radio telescopes. In Section~\ref{sec:gammaradio}, we study the $\gamma$-ray and radio flaring activity of J1048+7143. In Section~\ref{sec:predictFlare}, we model its $\gamma$-ray light curve as result of the spin-orbit precession of the dominant jet, predict the gravitational lifetime of the system and the next $\gamma$-ray flare. In Section~\ref{sec:gwrad}, we calculate predictions for the observation of the gravitational wave emission of this AGN. In Section~\ref{sec:sumconcl}, we summarize the results and give our conclusions.

Assuming the cosmological model with $H_0=69.6$~km\,s$^{-1}$\,Mpc$^{-1}$, $\Omega_\mathrm{M}=0.286$, and $\Omega_\Lambda=0.714$, the scale at the redshift of J1048+7143 $(z=1.15)$ is $8.361$~kpc\,arcsec$^{-1}$.

\section{$\gamma$-ray Light Curve and SED Analysis of J1048+7143}
\label{sec:gamma}
\subsection{Analysis of Fermi-LAT Data}
\begin{deluxetable*}{ccccccc}
\tablecaption{Exponential fitting of the $\gamma$-ray light curve of 4FGL~J1048.4+7143. Column (1) sets the fitted quantity ($a$ measures the height of the respective flare, $b$ the time location of its peak, and $c$ its slope). The numbered columns (2)--(7) contain the values of the parameters of the exponential functions fitted to the respective flares, where $F_{i,j}$ means the $j$-th subflare ($j$=1 or $j$=2) of the $i$-th main flare (from $1$ to $3$). The fitted baseline is $a_0=(0.14\pm0.03) \times 10^{-7} \mathrm{ph}~\mathrm{cm}^{-2} \mathrm{s}^{-1}$. 
\label{table:exp_flares}}
\tablewidth{0pt}
\tablehead{
\colhead{Parameter} & \colhead{$F_{1,1}$} & \colhead{$F_{1,2}$} & \colhead{$F_{2,1}$} & \colhead{$F_{2,2}$} & \colhead{$F_{3,1}$} & \colhead{$F_{3,2}$}}
\decimalcolnumbers
\startdata
$a$ ($\times10^{-7}~\mathrm{ph}~\mathrm{cm}^{-2} \mathrm{s}^{-1}$) & $1.87\pm0.40$ & $3.36\pm0.64$ & $2.60\pm0.70$ & $3.97\pm0.60$ & $4.01\pm0.58$ & $3.63\pm0.62$\\
$b$ (MJD, days) & $56378\pm17$ & $56710\pm10$ & $57578\pm15$ & $57801\pm10$ & $58760\pm8$ & $58957\pm7$ \\
$c$ (days) & $109\pm32$ & $70\pm17$ & $75\pm23$ & $87\pm17$ & $72\pm14$ & $58\pm13$ \\
\enddata
\end{deluxetable*}

\begin{figure*}
    \centering
    \includegraphics[angle=0,scale=0.6]{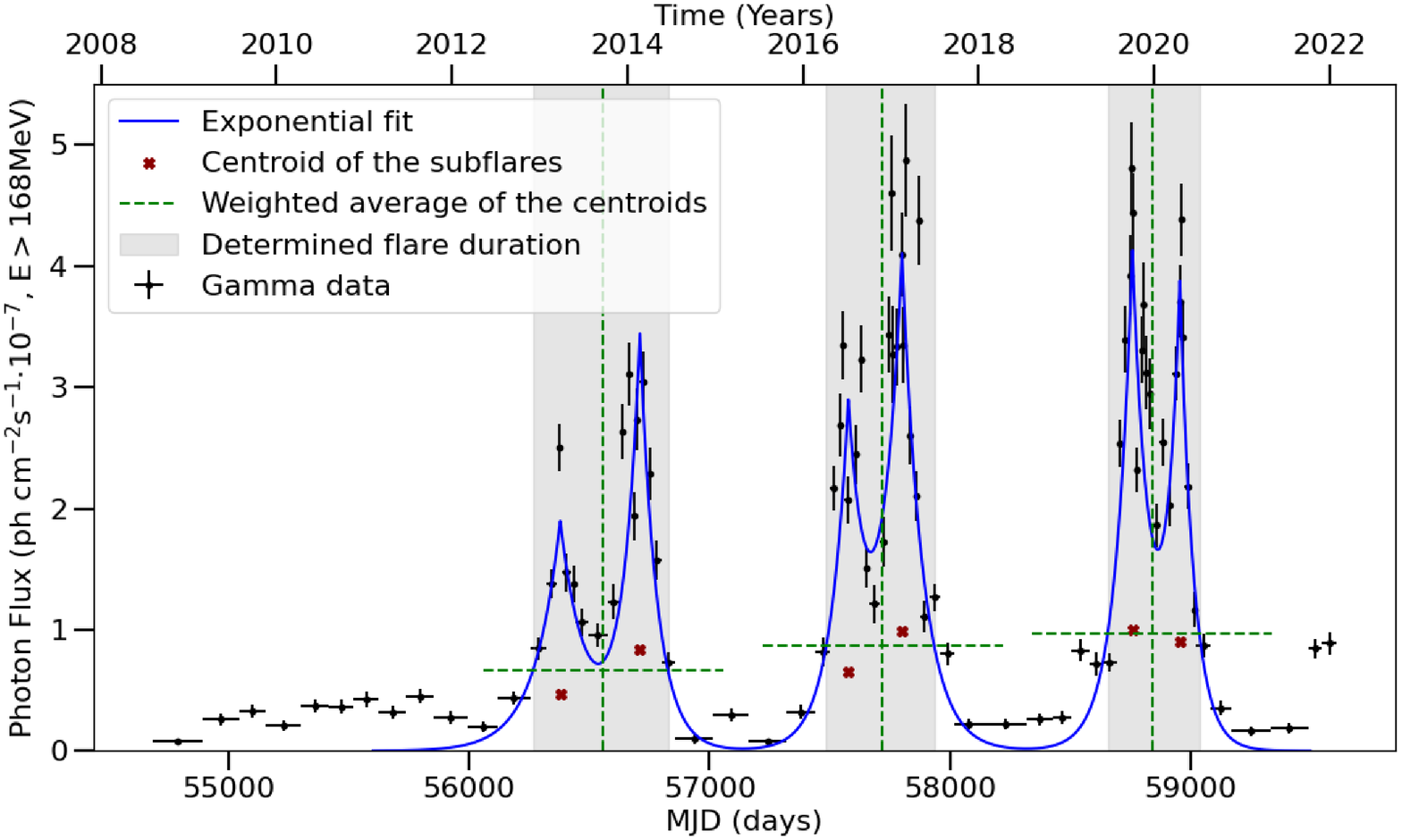}
    \caption{Illustration of the centroid method applied on the $\gamma$-ray flux curve of J1048+7143. The sum of the contributions of the exponential flares fitted to the light curve is shown by a blue continuous line. Centers of the major flares ($E>168$~MeV) are shown as green vertical dashed lines. The baseline is not shown as it has no impact on the duration or center of the flares.}
    \label{fig:centroid_Method}
\end{figure*}

To measure the $\gamma$-ray flux of 4FGL~J1048.4+7143 (J1048+7143), we obtained archival data taken with the LAT instrument onboard the Fermi Gamma-ray Space Telescope. This bright $\gamma$-ray source, with equatorial coordinates in the LAT 10-year Source Catalog \citep[4FGL,][]{Fermi4fgldr12020} Data Release 2 (DR2) right ascension $\mathrm{RA_{J2000}}=162\fdg1067$ and declination $\mathrm{DEC_{J2000}}=71\fdg7297$), is associated with the quasar J1048+7143. The region of interest (ROI) with a $15\degr$ radius was centered at the J2000 sky coordinates of 4FGL~J1048.4+7143. We collected almost 14 years of Fermi-LAT data in the time range 2008 Aug 4 -- 2022 Mar 14 (MJD 54682--59635), and in the energy range $100\,\mathrm{MeV}-800\,\mathrm{GeV}$.

We performed the unbinned likelihood analysis of the data utilizing the fermipy v1.0.1 and ScienceTools v2.0.8 packages, both built in the FermiBottle Docker container and analysis environment provided by the Fermi Science Support Center (FSSC)\footnote{\url{https://github.com/fermi-lat/FermiBottle}}. The analysis was carried out on a cluster of $16$ vCPU-s (Intel Skylake 16 $\times$ $2.2$ GHz) of the ELKH Cloud\footnote{\url{https://science-cloud.hu}}. We selected event type ``front+back'' (evtype=3) which is the recommended type for a point source analysis. The instrument response  function \verb+P8R3_SOURCE_V2+ was employed altogether with templates of the Galactic interstellar emission model \verb+gll_iem_v07.fits+ and of the isotropic diffuse emission \verb+iso_P8R3_SOURCE_V2_v1.txt+\footnote{\url{https://fermi.gsfc.nasa.gov/ssc/data/access/lat/BackgroundModels.html}}. We applied the nominal data quality cut \verb+(DATA_QUAL > 0) && (LAT_CONFIG==1)+, and a zenith angle cut $\theta<90\degr$ to eliminate Earth limb events. Time intervals when the Sun was closer than $15\degr$ to 4FGL~J1048.4+7143 were filtered out. Additional sources that were not in 4FGL-DR2 but were detected in our analysis have power-law flux profiles. We list the five additional sources we found in the Appendix in Table \ref{table:addsources}. The minimum separation allowed between two new individual point sources in the ROI was $0.3\degr$.

\subsection{Light Curve of J1048+7143 above 168 MeV}

While \citet{Wang2022} employed fixed bin width to construct the $\gamma$-ray light curve of 4FGL~J1048.4+7143, we applied the adaptive-binning algorithm of \citet{Lott2012} to set the bin widths of the light curve, in order to get more information about the flaring activity of this AGN. We tested 15\%, 7\%, 5\%, 1\% values of the relative flux uncertainties to adaptively set the bin widths of the light curve of 4FGL~J1048.4+7143. Based on the performances of the tests, we set the relative flux uncertainty to $5$\%. With smaller relative errors, we would lose information about the variability, while with larger relative errors the bin widths would become smaller and the light curve would be somewhat oversampled. With the adaptive binning method, we compute light curves for an integral flux above the optimum energy ($E_1$) for which the accumulation times are the shortest compared to other choices of a minimum energy. With this energy cut at 168 MeV, we also avoid the accidental capture of photons from the Galactic and extragalactic backgrounds, which strongly dominate over the emission of 4FGL~J1048.4+7143 at about $100$~MeV.

We present the photon flux of the bright $\gamma$-ray source 4FGL~J1048.4+7143 as function of time in Fig.~\ref{fig:gammalc}. From the beginning of the Fermi-LAT observations till about MJD 56000, the source was in a quiescent phase. After that, the source has shown three major flares, such that each of these flares can be separated into two subflares. We applied the method of the Lomb--Scargle periodogram \citep{Lomb1976,Scargle1982} on the light curve of J1048+7143, and found a period of $\sim 3.05$ yrs at a significance level of $5.5\sigma$. With this we confirm the $\sim 3.05\pm0.43$ yrs period found by \cite{Wang2022} at an even higher significance compared to their results ($\sim3.6\sigma$). We were able to achieve higher significance because we employed the adaptive binning method to derive the light curve bins, which makes our analysis even more sensitive to find light variations.
We fitted the photon flux of the major flares with two-sided exponential functions in the form of
\begin{equation}
    F_{i,j}(t)=a_0 + a_{i,j} \times \exp\left[{\frac{-|t-b_{i,j}|}{c_{i,j}}}\right],
\end{equation}
where $t$ is time, the index $i$ runs from $1$ to $3$ for the three main flares, $j$ is 1 or 2 for the subflares of the $i$th main flares, $a_0$ measures the baseline, $a_{i,j}$ the height, $b_{i,j}$ the time location of the peak, and $c_{i,j}$ the slope of the exponential function. Since in the quiescent phase the $\gamma$-ray flux did not vanish, we also fitted a constant baseline. We present the resulting fit parameters in Table~\ref{table:exp_flares}. The reduced $\chi^2$ of the fit is $11.2$.

Increasing the number of subflares for each main flare to 3 improves the fit ($\chi^2_\mathrm{red}=9.6$). However, in the present paper we only need the centroids of the major flares, which are only slightly affected if we assume a more complex structure of the major flares, and there is no physical reason to increase the number of subflares in our model. The substructure of the major flares will be addressed later.
 
 \begin{deluxetable*}{cccccc}
\tablecaption{Flare characteristics using the centroid method. 
\label{table:flare_characteristics}}
\tablewidth{0pt}
\tablehead{
\colhead{par} & \colhead{F1} & \colhead{$P_{1\to2}$} & \colhead{F2} & \colhead{$P_{2\to3}$} & \colhead{F3}}
\startdata
Flare center (MJD) & $56556 \pm 69$  && $57720 \pm 53$  && $58843 \pm 44$ \\
Flare duration (days) & $561 \pm 81$  && $449 \pm 89$  && $383 \pm 57$  \\
\multicolumn{1}{p{2.8cm}}{\centering Time till next flare center (years)} & & \multirow{2}{*}{$3.19 \pm 0.24$} && \multirow{2}{*}{$3.07 \pm 0.19$} & \\
\enddata
\end{deluxetable*}

\subsection{Time Characteristics of the $\gamma$-ray Flares with the Centroid Method}
\label{centroid}

In order to determine the time duration between the major flares detected in the $\gamma$-ray flux, the duration of the flares is defined as a clear delimitation of where a flare starts and where it ends. For that, we determine the geometric center (centroid) of each of the six subflares based on their mathematical surface integral (red crosses in Fig.~\ref{fig:centroid_Method}). This is the so-called centroid method. 
The $x$(=time) and $y$(=photon flux) coordinates of the geometric center ($X_{i,j},Y_{i,j}$) of the $i,j$th subflare are calculated as
\begin{equation}
    X_{i,j} = \frac{\int t \cdot F_{i,j}(t) \, \mathrm{d}t}{\int F_{i,j}(t) \, \mathrm{d}t}, 
\end{equation}
\begin{equation}
    Y_{i,j} = \frac{1}{2} \frac{\int F_{i,j}^2(t) \, \mathrm{d}t}{\int F_{i,j}(t) \, \mathrm{d}t},
\end{equation}
respectively. The weighted average of the subflares from one major flare yields the center of the latter, indicated as the cross of the dotted green lines. 
If $A_i$ and $B_i$ are the surface integrals below the exponential fit to the first and second subflare of the $i$th main flare, then the $i$th vertical dotted green line crosses the $x$-axis at
\begin{eqnarray}
    X_i=\frac{A_{i}}{A_{i}+B_{i}} X_{(i,1)}+\frac{B_{i}}{A_{i}+B_{i}} X_{(i,2)},
\end{eqnarray}
and the height of the $i$th horizontal dotted green line is calculated as
\begin{eqnarray}
    Y_i=\frac{A_{i}}{A_{i}+B_{i}} Y_{(i,1)}+\frac{B_{i}}{A_{i}+B_{i}} Y_{(i,2)}.
\end{eqnarray}
Then the intersection of the $i$th horizontal and vertical dotted green lines is at $X_i$, $Y_i$. The horizontal intersections of the dotted green lines through the center with the exponential fit mark the beginning and end of the major flare.
The inferred flare durations are marked as gray areas with their errors in Fig.~\ref{fig:centroid_Method} and are given in Table \ref{table:flare_characteristics}.
The uncertainty in the durations is computed using the maximum error of the fitting results obtained for each subflare (see Table~\ref{table:exp_flares}).
We defined the period as the elapsed time between the centers of corresponding major flares and found for the period between major flare one and two, $P_{1\to2} = 3.19 \pm 0.24$~yr and for the second period between the major flares two and three: $P_{2\to3} = 3.07 \pm 0.19$~yr (see Table~\ref{table:flare_characteristics}).

\subsection{Gamma SED}

\begin{figure}
    \centering
    \includegraphics[scale=0.6]{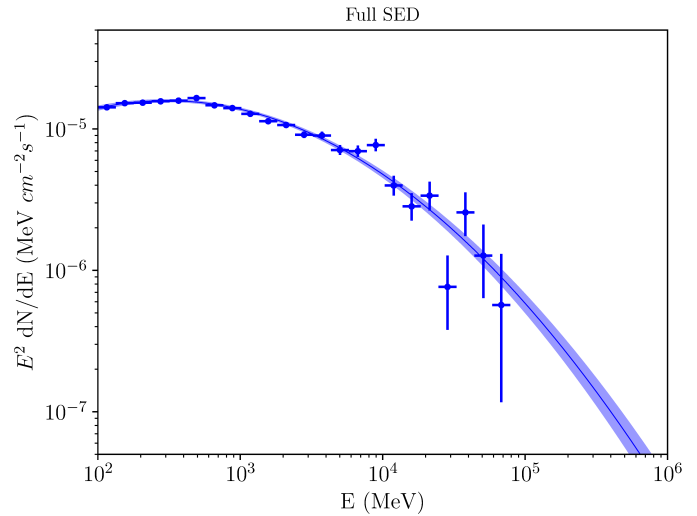}
    \caption{The spectral energy distribution (SED) of 4FGL~J1048.4+7143 in the full Fermi-LAT time range.}
    \label{fig:seds}
\end{figure}

The spectral energy distribution ($E^2 dN/dE$) of the source was derived from the differential flux which is described by a logparabola shape as:
\begin{equation}
\frac{dN}{dE}=\phi_0 \left(\frac{E}{E_0}\right)^{-\alpha-\beta \ln{E/E_0}},
\end{equation}
where $\phi_0$ is the prefactor, $\alpha$ is the index, $\beta$ is the curvature parameter, and $E_0$ is the scale parameter. The SED of 4FGL~J1048.4+7143 generated in the full time range (2008 Aug 4 -- 2022 Mar 14, MJD 54682--59635) extends up to $\sim 100$~GeV, with best-fit prefactor $\phi_0=(3.00\pm0.04)\times 10^{-11} \mathrm{MeV}\,\mathrm{cm}^{-2}\,\mathrm{s}^{-1}$, index $\alpha=2.16\pm0.01$, curvature parameter $\beta=0.10\pm0.01$, scale factor $E_0=702.37$~MeV, the latter held fixed at its 4FGL-DR2 value). We present the SED in Fig.~\ref{fig:seds}.

\begin{figure*}
\includegraphics[scale=0.27]{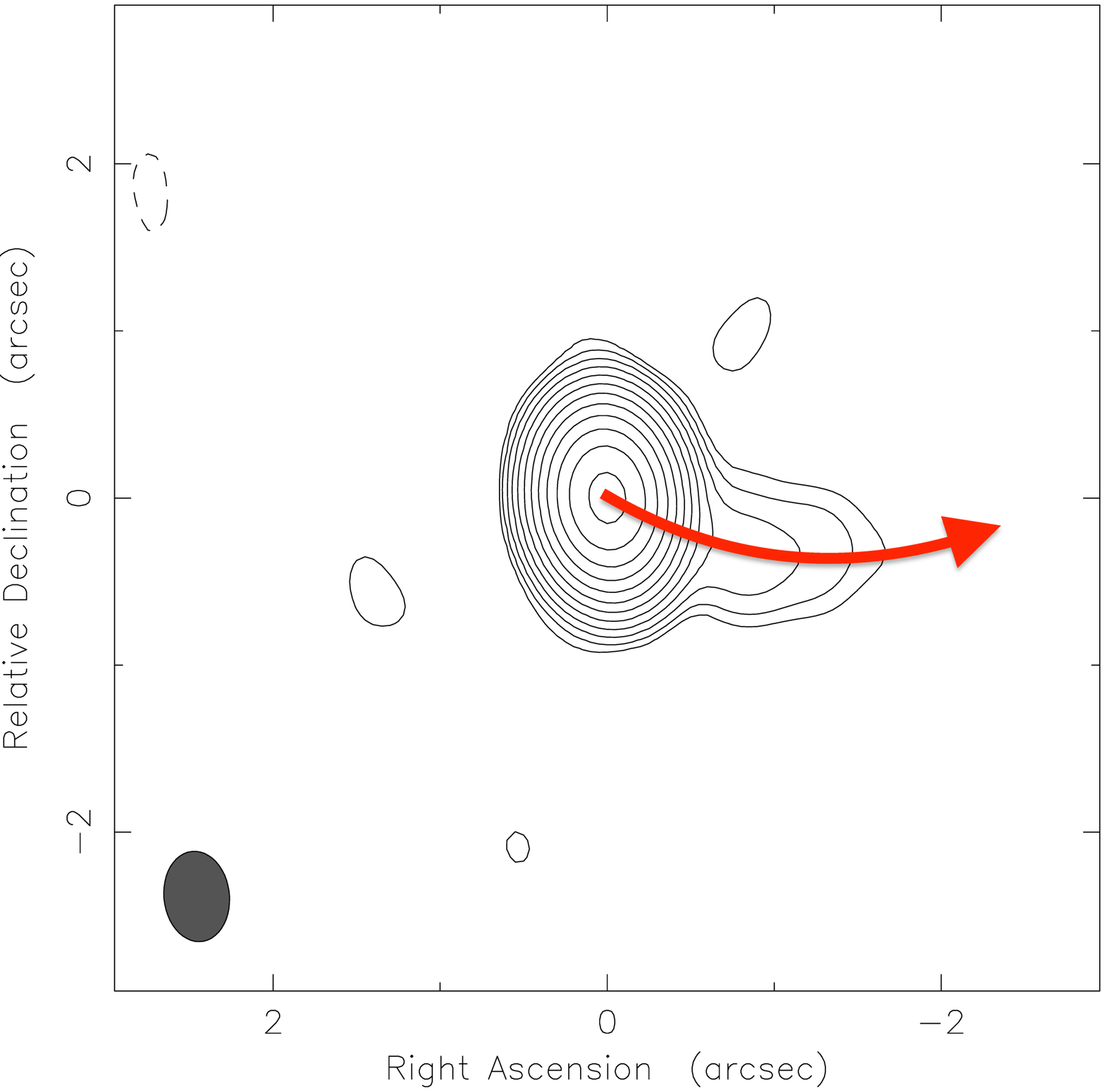}
\includegraphics[angle=90,scale=0.27]{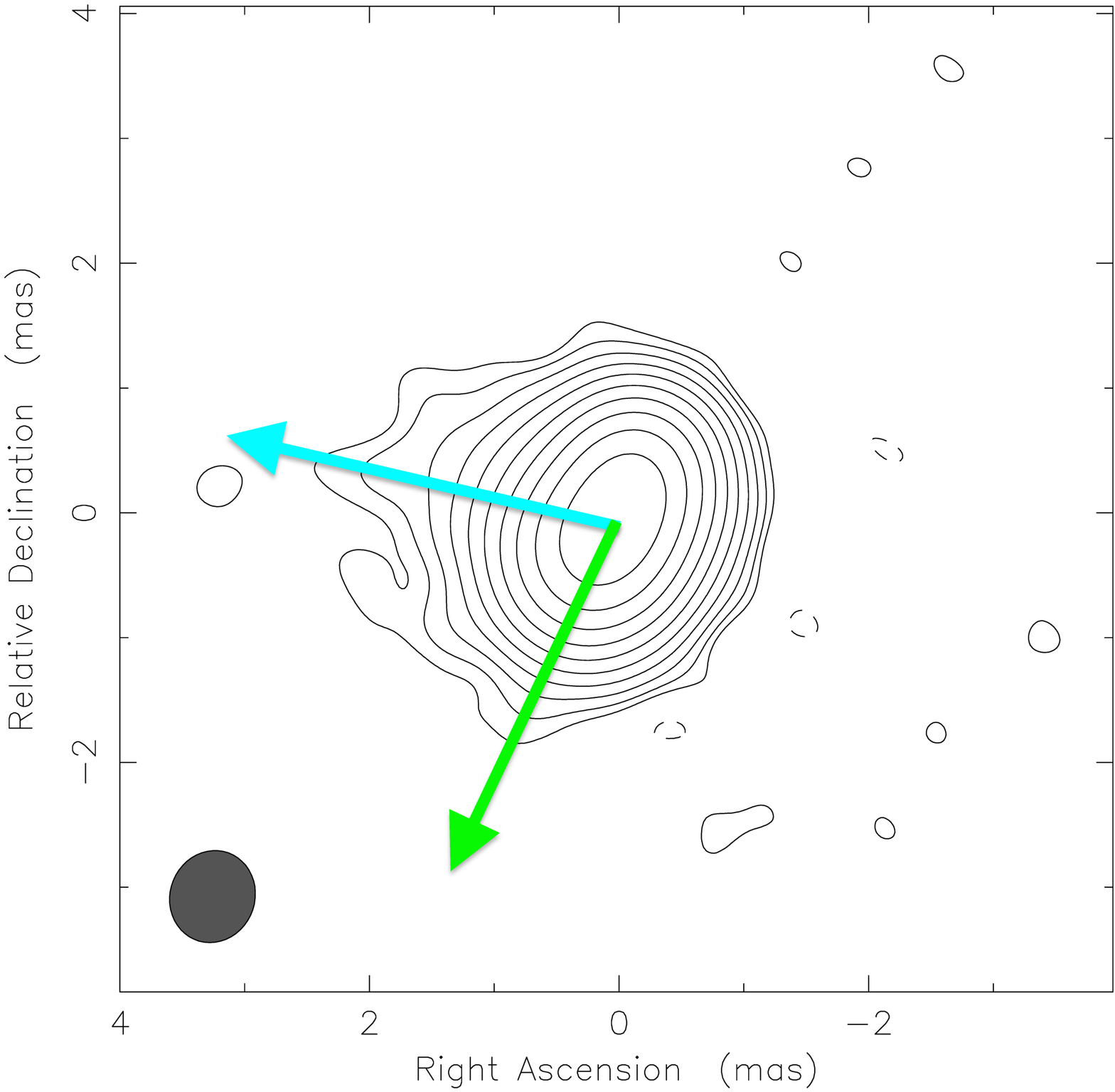}
\includegraphics[scale=0.27]{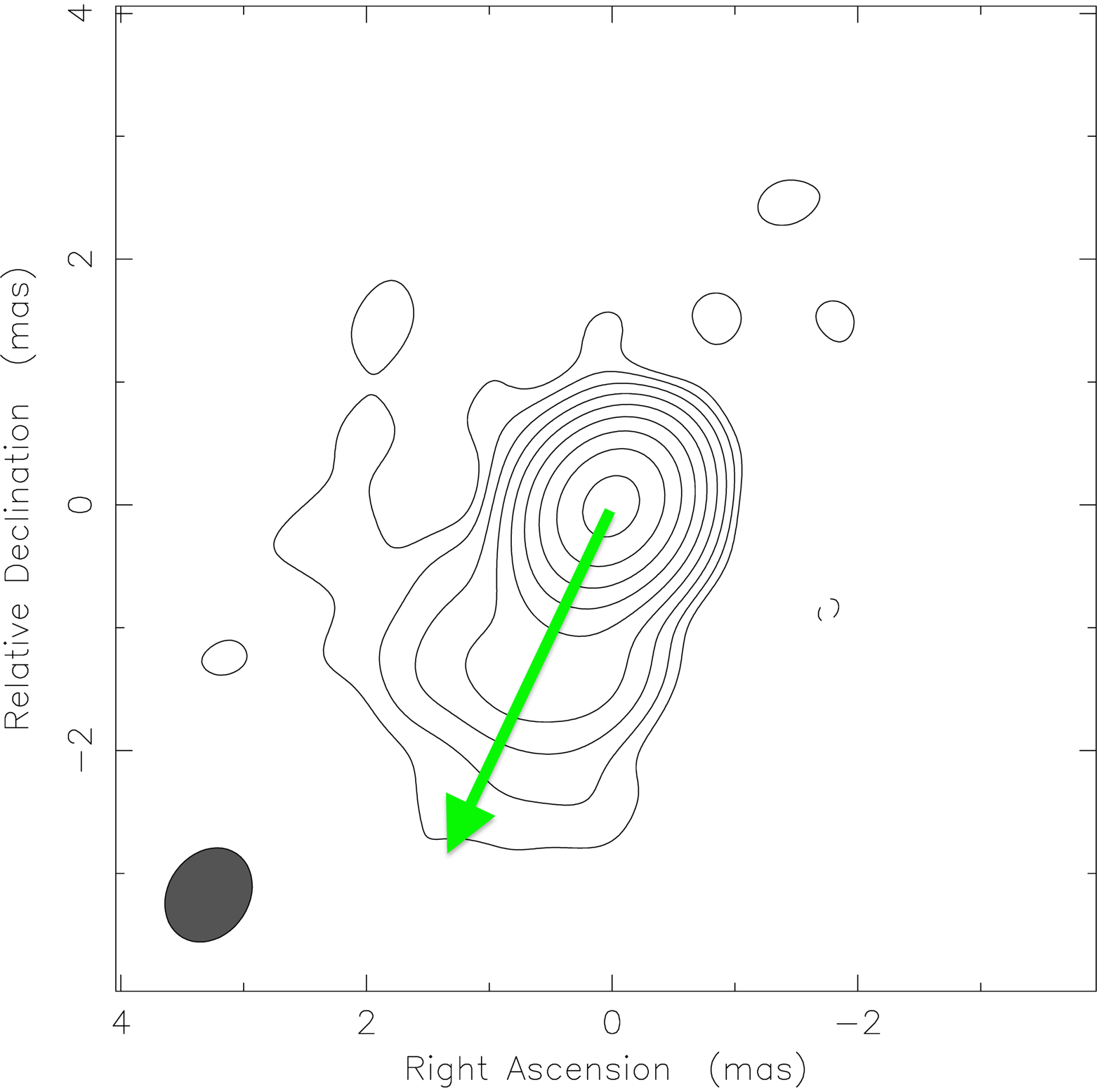}
\caption{\textit{Left:} The $4.8$-GHz VLA image of J1048+7143 with the jet oriented towards west and extending to $\sim 15$~kpc projected distance (red arrow). The peak brightness is $1.66$~Jy\,beam$^{-1}$, the lowest contours are at $\pm 0.654$~mJy\,beam$^{-1}$ ($\sim 3\sigma)$, the positive contours increase by a factor of $2$. The elliptical Gaussian restoring beam is $0\farcs542 \times 0\farcs392$ with major axis position angle $\mathrm{PA}=6.2\degr$. \textit{Middle and right:} Two examples of the $8.6$-GHz jet structure observed with the VLBI. The angular scale of $1$~mas corresponds to $8.361$~pc projected linear scale. In the mas-scale structure, there is an approximately eastern (turquoise arrow) and a southern jet extension (green arrow), such that sometimes the eastern and sometimes the southern jet dominates, as discussed in the text. The image parameters in the middle are: observing date 2001 Jul 5, peak brightness 759~mJy\,beam$^{-1}$, lowest contours $\pm 0.76$~mJy\,beam$^{-1}$ (the positive contours
increase by a factor of $2$), restoring beam $0.75\,\mathrm{mas} \times 0.68\,\mathrm{mas}$ at $\mathrm{PA}=-20.3\degr$. On the right: observing date 2010 Mar 23, peak brightness 882~mJy\,beam$^{-1}$, lowest contours $\pm 2.6$~mJy\,beam$^{-1}$ (the positive contours
increase by a factor of $2$), restoring beam $0.81\,\mathrm{mas} \times 0.66\,\mathrm{mas}$ at $\mathrm{PA}=34.4\degr$.
\label{fig:maps}}
\end{figure*}

\section{Kpc- and pc-scale Radio Structure of J1048+7143, Signs of Jet Precession}
\label{sec:radio}

\subsection{Analysis of Archival VLA Observations}

Data from several VLA imaging experiments targeting J1048+7143 are available in the U.S. National Radio Astronomy Observatory (NRAO) archive\footnote{\url{https://data.nrao.edu/}}. We chose experiment AG512 (PI: L. Greenhill) conducted on 1997 Jan 12. The quasar was used as a calibrator source for observing the water megamaser galaxy NGC\,3735 \citep{Greenhill1997}
at multiple frequencies ($4.8$-, $8.4$-, and $15$-GHz). The array was in its most extended A configuration. We calibrated the data in the NRAO Astronomical Image Processing System \citep[AIPS,][]{Greisen2003} in a standard way.
The source 3C\,286 was used as the primary flux density calibrator. To indicate the kpc-scale structure, here we present the $4.8$-GHz image of J1048+7143 (Fig.~\ref{fig:maps}, left). It was made with the software \textsc{Difmap} \citep{Shepherd1997} using the calibrated visibility data exported from AIPS. The total on-source observing time was $500$~s, the bandwidth $100$~MHz. The $8.4$-GHz image showed a qualitatively similar structure with a weak jet pointing to the western direction, extending to $\sim 15$~kpc projected linear size, while the weak extended jet emission was resolved out and remained undetected at $15$~GHz. Notably, a published VLA image made at a lower frequency, $1.4$~GHz \citep{Xu1995}, indicates in addition a weak feature at $\sim 13\arcsec$ towards the north-northwest, in a position angle misaligned by almost $90\degr$ with respect to the arcsec-scale jet seen in Fig.~\ref{fig:maps}.

\subsection{Analysis of Archival VLBI Radio Observations}

To derive the structural and kinematic properties of the mas-scale jet of J1048+7143, we used archival calibrated $8.6$-GHz visibility data taken with the VLBA, occasionally supplemented with other radio telescopes to form a global VLBI network. The observations span more than $26$~yr at 69 epochs (between 1994.61 and 2020.73), and the calibrated visibilities are available in the Astrogeo database\footnote{\url{http://astrogeo.org/cgi-bin/imdb_get_source.csh?source=J1048\%2B7143}}. 
The original observations were made in the framework of various projects, e.g. the VLBA Calibrator Survey \citep{Beasley2002}, astrometric monitoring programs of radio reference frame sources \citep[e.g.][]{Fey1997,Petrov2009,Petrov2011,Petrov2013,Shu2017,Petrov2021}, and imaging Fermi $\gamma$-ray sources \citep{Schinzel2015,Schinzel2017}.
Our images were made with \textsc {Difmap} \citep{Shepherd1997}. The brightness distribution of the source was modeled using the visibility data \citep{Pearson1995} by fitting elliptical Gaussian components to the core and circular Gaussians to the jet components. The fitted parameters are the component flux density, the position, and the full width at half-maximum (FWHM) diameter of the circular components, or the major axis and the minor-to-major axial ratio for the ellipticals. We followed \citet{Kun2014} for the error estimation of the fitted parameters. Two examples of the $8.6$-GHz VLBI images are shown in Fig.~\ref{fig:maps} (middle and right panels). 
\begin{figure*}
    \centering
    \includegraphics[angle=270,scale=0.8]{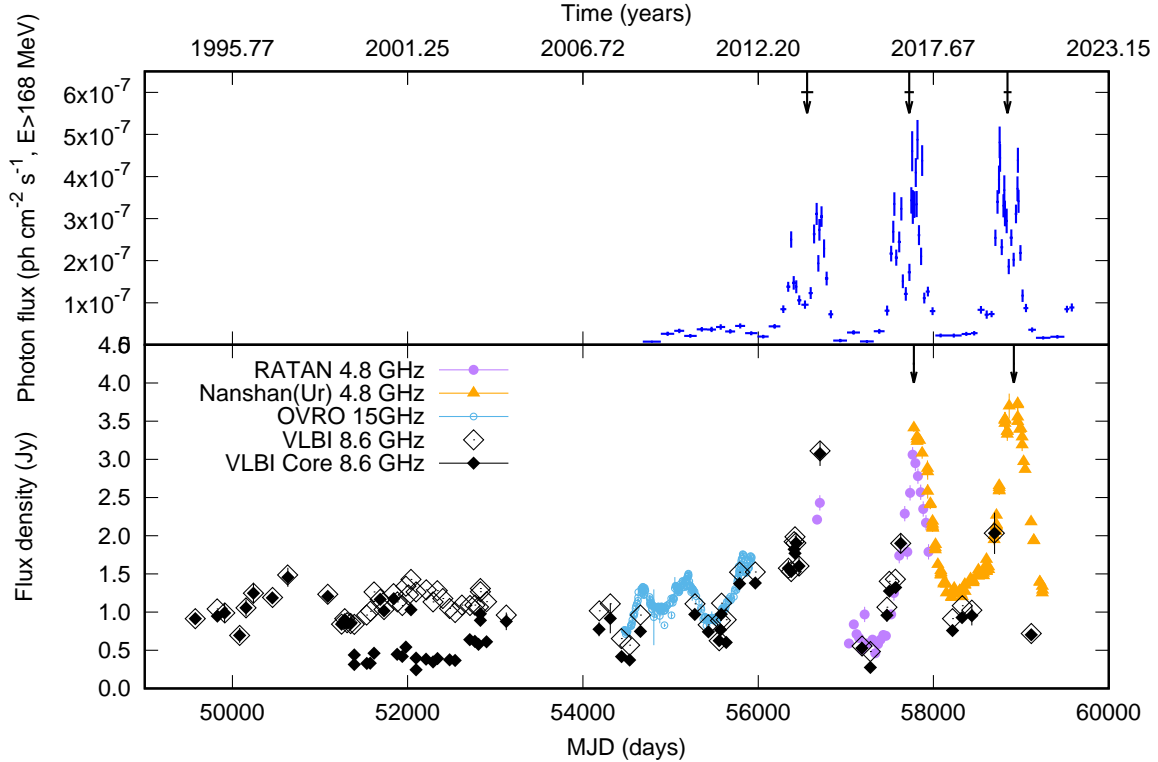}
    \caption{Upper panel: the Fermi-LAT photon flux ($186\,\mathrm{MeV} < E < 800\,\mathrm{GeV}$) of 4FGL~J1048.4+7143 as a function of time. Lower panel: the RATAN-600 $4.8$-GHz (purple filled circles), Nanshan $4.8$-GHz (orange triangles), OVRO $15$-GHz single-dish radio flux density curves (blue empty diamonds), the $8.6$-GHz integrated interferometric flux density of J1048+7143 (black empty diamonds), and the $8.6$-GHz flux density of the VLBI core (black filled diamonds).}
    \label{fig:radio_flux_curves}
\end{figure*}
\subsection{Jet Structure of J1048+7143 at pc Scales}

We note that the coverage of the $(u,v)$ plane is much denser for the observations in the beginning of 2000's, compared to the rest of the observations, which results in better sensitivity. In the middle panel of Fig.~\ref{fig:maps} (observing date 2001 Jul 5), one can see that there is a sign of a jet pointing approximately toward east and one pointing to south. In the right panel of the same figure (observing date 2010 Mar 23), at another epoch, one can see a south-directed jet, while there is no conclusive indication of the east-directed jet. Also, the large misalignment between the arcsec-scale and mas-scale jets suggest jet precession \citep{Conway1993}. We will investigate the jet kinematics based on the well-defined jet components of the $8.6$-GHz VLBI structure of J1048+7143 in a subsequent paper, while here we focus on the flaring activity of the source. 

\subsection{Single-dish Radio Observations at $4.8$-GHz}
\label{sec:singledish}
A long-term monthly flux density monitoring program for a sample of about 100 $\gamma$-ray AGN at $4.8$- and $23$-GHz frequencies using the Nanshan 26-m radio telescope has been conducted in the Xinjiang Astronomical Observatory (Urumqi, China) since early 2017. Our target source, J1048+7143, is part of the monitoring program. This source was observed between 2017 Jan 24 and 2021 Jan 29 in a total of 102 epochs. The data to derive the single-dish flux densities were obtained in dual circular polarization with the central frequency of $4.8$-GHz and bandwidth of 600~MHz. The typical system temperature is 24~K in clear weather, and the antenna sensitivity is $\sim0.12$~K\,Jy$^{-1}$ \citep[e.g.][]{Liu2015}.

J1048+7143 was also observed with the 600-m RATAN-600 radio telescope in Russia, in 28 epochs between 2014 Jan 15 and 2017 Jul 10 \citep{Mingaliev2014}. We obtained public data taken at $4.8$-GHz from the RATAN-600 multi-frequency catalog of BL Lac objects\footnote{\url{http://www.sao.ru/blcat/}}. We show the Nanshan and RATAN-600 flux density curves in Fig.~\ref{fig:radio_flux_curves}. While both light curves were obtained at the same frequency, it is apparent from Fig.~\ref{fig:radio_flux_curves} that the RATAN-600 values are systematically below the Nanshan flux densities in an overlapping period in early 2017. Assuming proper calibrations, the flux density difference in the order of $\sim0.1$~Jy most probably is caused by some extended emission resolved out with the large 600-m diameter antenna but detected with the $>20$ times smaller dish in Nanshan. For the purposes of this study, the trends in the flux density variability are important, and the rapidly declining flux densities in early 2017 are perfectly consistent in both radio monitoring data sets (Fig.~\ref{fig:radio_flux_curves}).

\section{Comparison of the $\gamma$-ray and the Radio Flaring Activity of J1048+7143}
\label{sec:gammaradio}

We present the Fermi-LAT photon flux at the top of Fig.~\ref{fig:radio_flux_curves}, along with the Nanshan and RATAN-600 radio flux density curves,  the integrated and core $8.6$-GHz flux densities obtained from VLBI observations in the bottom panel of Fig.~\ref{fig:radio_flux_curves}.
The first Fermi-LAT bin is centered to 2008 Nov 15 23:52:48 UTC and its width is about 206 days. The flaring activity of J1048+7143 in the $\gamma$-ray regime ($E>168$~MeV) started in the Spring of 2012. In Section \ref{centroid}, we determined the center of three main $\gamma$-flares as $T_{0,\mathrm{F1}}=\mathrm{MJD}~56556\pm69$ days, $T_{0,\mathrm{F2}}=\mathrm{MJD}~57720\pm53$ days, $T_{0,\mathrm{F3}}=\mathrm{MJD}~58843\pm44$ days, respectively. We mark the centers of the main $\gamma$-flares with down-pointing arrows in the upper panel of Fig.~\ref{fig:radio_flux_curves}.

We augmented the interferometric radio data with public $15$-GHz single dish data taken with the 40-m radio telescope of the Owens Valley Radio Observatory \citep[OVRO,][]{Richards2014}. Before 2008, although the source shows radio variability, we do not see large radio flares at either frequency bands ($\sim 5$ and $8$-GHz). The flaring activity of J1048+7143 in the radio regime (single-dish $4.8$-GHz, interferometric $8.6$-GHz) started also in the Spring of 2012, simultaneously with the $\gamma$-ray regime. We fitted two-sided exponential functions to the radio flux density curve of J1048+7143, where, to correct for the resolution effect (see Section \ref{sec:singledish}.), we shifted the RATAN flux density curve to the Nanshan (Ur) curve along the flux density axis. With this, we determined the centers of the two main radio flares as $T_{0,\mathrm{RF1}}=\mathrm{MJD}~57804\pm7$ days, $T_{0,\mathrm{RF2}}=\mathrm{MJD}~58912\pm4$ days, respectively. We mark the center of the main radio flares with down-pointing arrows in the lower panel of Fig.~\ref{fig:radio_flux_curves}. Although we do not have Fermi-LAT observations from earlier than 2008, the timing of the $\gamma$-ray and radio activity of the source seems to be parallel in the time range covered by Fermi-LAT observations. From the numbers, we can see that the two complete radio flares slightly lag behind the $\gamma$-ray flares, with $84\pm54$ and $69\pm44$ days, respectively.

In Section \ref{centroid}, we found the time elapsed between the first and second major $\gamma$-flares as $P_{1\to2} = 3.19 \pm 0.24$~yr and the time elapsed between the second and third major $\gamma$-flares as $P_{2\to3} = 3.07 \pm 0.19$~yr (see Table~\ref{table:flare_characteristics}). We compared these time durations between the $\gamma$-ray flares with the ones inferred from the radio flux density (see Fig.~\ref{fig:radio_flux_curves}). For the latter, we used the $8.6$-GHz VLBI data point at around 56700 MJD as a rough estimate of the first radio flare center and determined the other two radio flare centers with the FWHM of the RATAN-600 and Nanshan $4.8$-GHz total flux density data. 
We obtained an approximate value of $P_{\mathrm{R}, 1\to2}=3.06 \pm 0.03$~yr for the first time duration in radio and $P_{\mathrm{R}, 2\to3}=2.98 \pm 0.03$~yr for the second time duration in radio. 
Radio time durations $P_{\mathrm{R}, 1\to2}$ and $P_{\mathrm{R}, 2\to3}$ are inside the error range of the $\gamma$-ray time durations, indicating a possible connection between the two wavebands. 

During the flaring activity, the radio emission of J1048+7143 is mostly dominated by the compact VLBI core (see Fig.~\ref{fig:radio_flux_curves}). It means that with single-dish observations we mainly detect the radio emission of the VLBI core. Based on Gaussian fits at 69 epochs of $8.6$-GHz VLBI data, the average, standard deviation, median, minimum, and maximum of the major axis of the $8.6$-GHz VLBI core are $0.46$~mas (corresponding to projected linear size $3.96$~pc), $0.29$~mas ($2.50$~pc), $0.39$~mas ($3.37$~pc), $0.10$~mas ($0.84$~pc), and $1.81$~mas ($15.69$~pc), respectively. The average axial ratio of the minor and major axis of the core is $0.43$ with a standard deviation of $0.25$ (median value $0.46$). We note that the $8.6$-GHz VLBI observations around 2001 were more sensitive and the core region was resolved into more components compared to the rest of the observations. Probably that is why we do not see that the core dominates the total flux density curve about between MJD 51314 (1999.4) and MJD 52880 (2003.6), as seen in Fig. \ref{fig:radio_flux_curves}.

If the $\gamma$-ray and radio flares are physically connected, as suggested by the light curves, then the $\gamma$-ray emission comes from the VLBI core, and the size of the emission site can be restricted to pc or rather sub-pc scales. This constrains the emission site to be located well below the narrow-line region ($100-1000$~pc) and even below the torus ($1-100$~pc) region, somewhere probably inside the broad-line region. The average time lag between the gamma and the radio flares emerged as $77\pm70$ days, which means $0.065\pm0.058$ pc spatial difference between the two emission sites, assuming the perturbation that leads to the flares propagates with the speed of the light.

Such flaring behavior can be realized in terms of a precessing jet, where the changing line of sight angle of the jet causes changing apparent flux density due to variations in the Doppler factor
\begin{eqnarray}
\delta(t)=\frac{1}{\gamma(1-\beta\cos \iota(t))},
\label{eq:dopplerfactor}
\end{eqnarray}
where $\gamma$ is the Lorentz factor, $\beta$ is the jet velocity expressed in the unit of the speed of light, and $\iota(t)$ is the time-dependent line-of-sight angle (inclination angle) of the jet. 

The evident difference between the flaring activity in the $\gamma$-ray light curve and the radio flux density curve is that while in $\gamma$-rays we see sharp sub-flares, in radio we only see major flares without substructure. We attempt to explain this in a subsequent paper, such that the flares in $\gamma$-rays are results of the interaction of two spine--sheath jets combined with periodic Doppler boosting, and the radio variation comes solely from the periodic Doppler boosting of the dominant jet. We qualitatively explain the non-flaring behavior of J1048+7143 before 2012 in the next Section, while we will discuss it quantitatively in a subsequent paper.

\section{Spin--orbit Precession and Prediction of the Next $\gamma$-ray Flare}
\label{sec:predictFlare}

\begin{figure*}
    \centering
    \includegraphics[angle=0,scale=0.6]{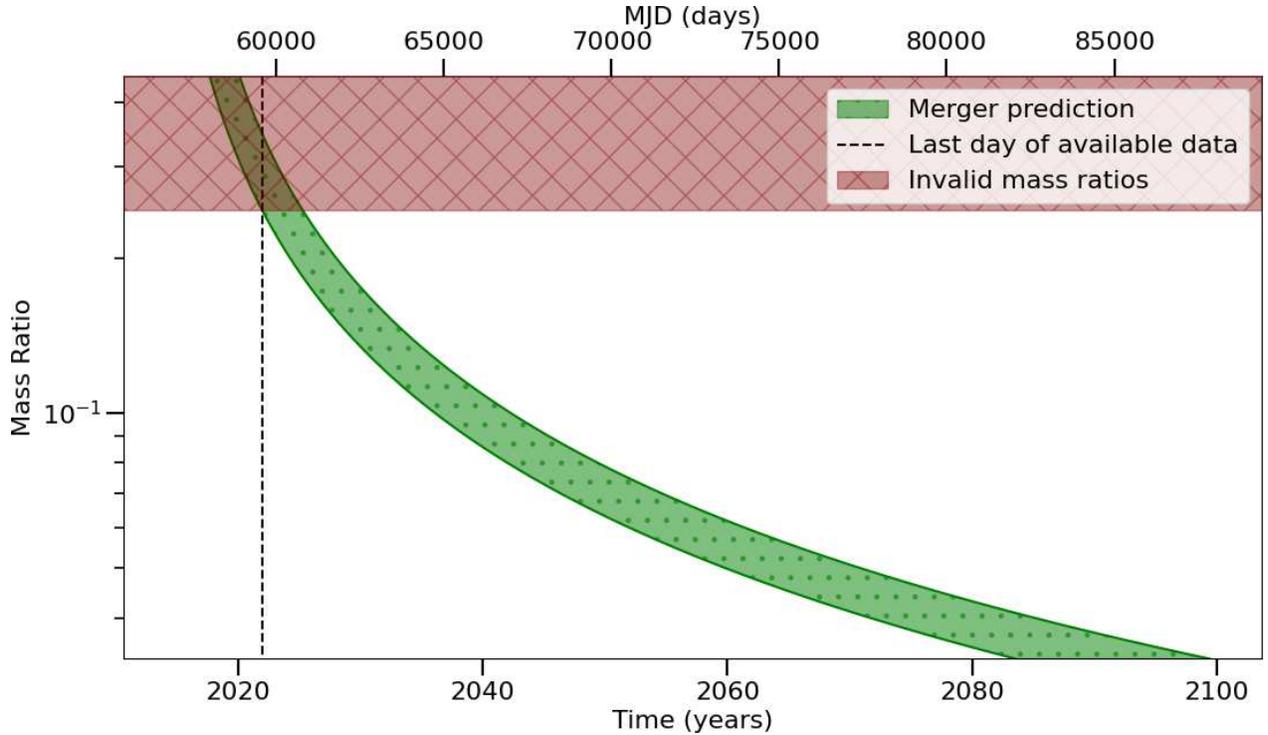}
    \caption{Date of the merger in case of a supermassive binary black hole at the center of J1048+7143, as a function of the mass ratio of the binary $q$.}
    \label{fig:mergerPredict}
\end{figure*}
\subsection{Jet Precession due to Spin--Orbit Precession}
\begin{figure*}
    \centering
    \includegraphics[angle=0,scale=0.6]{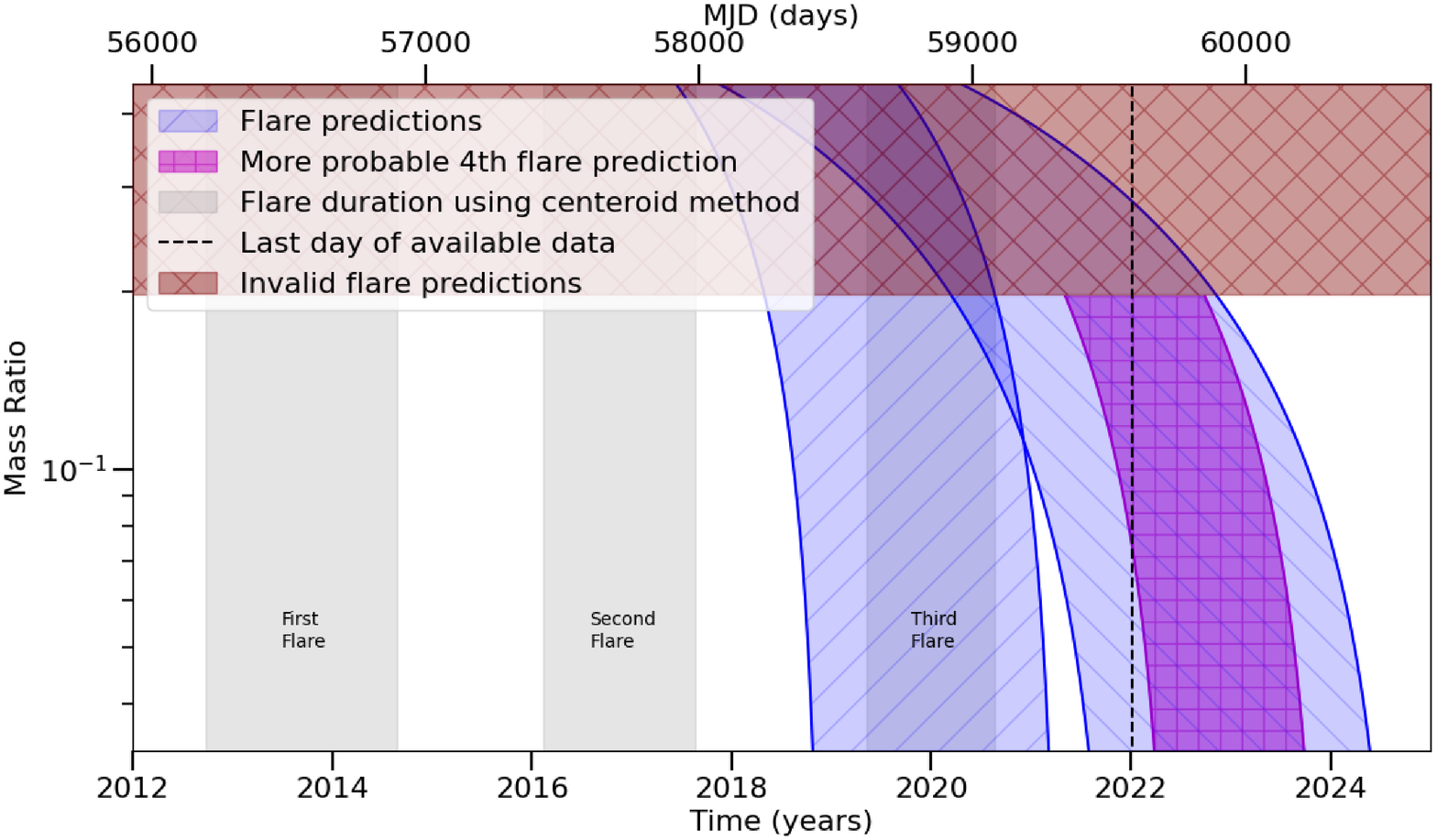}
    \caption{Prediction of the fourth flare from J1048+7143 in case of a SMBBH merger at its center. Red area: invalid flare prediction, as the merger would have already happened in the past. Gray area: duration of the first three flares detected with the errors in the duration. Blue area: flare predictions for the 3rd and 4th flare, respectively. The third flare prediction (diagonal lines from lower left to upper right) is consistent with the time at which the actual flare was detected. The fourth flare prediction (diagonal lines from lower right to upper left) contains the purple area, which highlights a more probable time span based on the actual duration of the third detected flare.}
    \label{fig:flarePredict}
\end{figure*}
Quasi-periodic flaring emissions, as seen in the $\gamma$-ray flux in Fig.~\ref{fig:gammalc}, are expected signatures from supermassive binary black holes, caused by the precession of the relativistic jet due to spin--orbit interactions close to the coalescence \citep[e.g.][]{Kun2020}. 
Therefore, we test for the (quasi-)periodic behavior of the three major flares, whether they are in agreement with an SMBBH merger at the core of J1048$+$7143.
We expand the jet precession model of \cite{deBruijn2020}, which determines the direction angle of dominant spin $\phi$ as a function of the remaining time until the binary coalescence, $\Delta T_{\rm GW}$. The angle $\phi$ changes from $0\degr$ to $360\degr$ in one spin-orbit precession period and it is measured in the plane perpendicular to the total angular momentum \citep[see e.g. in][]{deBruijn2020}. This model uses a 2.5 post-Newtonian (PN) order approximation (\citealt{GerPLB2009}). We modified the model of \cite{deBruijn2020} by allowing small mass ratios of the binary $q$, with $q \leq 1$, by multiplying by a factor ${(4 + 3q)}/(4 (1 + q)^2)$ and results in
\begin{eqnarray}
    &&\phi(\Delta T_{\rm GW} \,,\, q) = - \frac{2 \, (4 + 3q)}{(1 + q)^2}\times \nonumber \\
           &&\, \times \left( \frac{5 \, c}{32 \,G^{1/3} M^{1/3}} \cdot \frac{(1 + q)^2}{q} \right)^{3/4} \left(\Delta T_{\rm GW}\right)^{1/4} 
           + \psi \,.
           \label{eq:JetModel}
\end{eqnarray}
Here, $G$ is the gravitational constant, $c$ is the speed of light, $M$ is the total mass of the SMBBH, 
and $\psi$ is an integration constant, defining the initial direction of the jet in the inspiral phase of the merger before it changes significantly due to spin--orbit interactions. The spin--orbit precession formalism of \citet{GerPLB2009} assumes the secondary spin can be neglected, therefore the direction of the jet is assumed to be parallel with the dominant spin \citep[see the reasoning e.g. in][]{Kun2014}.

In our model, the quasi-periodic feature of the $\gamma$-ray light curve is caused by the quasi-periodic Doppler boosting of a precessing jet (see Eq.~\ref{eq:dopplerfactor}). Detecting two separate flares from one source, with a faint state in between, means in the jet-precession model that the jet made a full rotation and points at the observer again. Thus, we can establish the connection between the two flares:
\begin{equation}
    \phi(\Delta T_{\rm GW} \,,\, q) = \phi(\Delta T_{\rm GW} - P_{\rm jet} \,,\, q) \pm \zeta \,.
    \label{eq:determine_T_GW}
\end{equation}
Since some time has passed between the detection of the two flares, denoted as the precession period $P_{\rm jet}$, the time remaining until the binary coalescence decreased from $\Delta T_{\rm GW}$ to $\Delta T_{\rm GW} - P_{\rm jet}$ from the first to the second flare.
The parameter $\zeta$ captures the half-opening angle of the jet, which covers the duration of the flare.
Inserting Eq.~(\ref{eq:JetModel}), once as a function of $\Delta T_{\rm GW}$ and once as a function of $\Delta T_{\rm GW} - P_{\rm jet}$ into Eq.~(\ref{eq:determine_T_GW}) results in the elimination of the constant $\psi$ in Eq.~(\ref{eq:determine_T_GW}).

Subsequently, the time until the final coalescence of the SMBBH, $\Delta T_{\rm GW}$ (see Eq.~(\ref{eq:determine_T_GW})), can be determined as a function of the mass ratio $q$, if the total mass and the time between two flares, $P_{\rm jet}$, are given.

The quasi-periodic behavior only becomes visible starting in 2012, the lack of variations before 2012 is in need of explanation. Below, we outline how such a change from a rather constant flux to a quasi-periodic one happens within the frame of the jet precession model. As mentioned in \cite{GerPLB2009}, the total angular momentum $\mathbf{J}$ represents an invariant direction up to the 2nd PN order, the dominant spin $\mathbf{S}_1$ and the Newtonian angular momentum $\mathbf{L}$ obeying a precessional motion about $\mathbf{J}$. The dynamics becomes dissipative from 2.5PN order, the direction of $\mathbf{S_1}$, the magnitude and the direction of $\mathbf{L}$ also change due to the gravitational radiation of the SMBBH such that \citep{GerPLB2009}:
\begin{eqnarray}
    \mathbf{\dot{\hat{S}}_1}&=&\frac{2G}{c^2r^3} \mathbf{J}\times\mathbf{\hat{S}_1},\\
    \dot{L}&=&-\frac{32G\mu^2}{5r}\left( \frac{Gm}{c^2r}\right)^{5/2},\\
    \mathbf{\dot{\hat{L}}}&=&\frac{2G}{c^2 r^3} \mathbf{J}\times \mathbf{\hat{L}}.
\end{eqnarray}
The length of the spin vector does not change since the spin rotational energy is channelled into the jets only on the Hubble time scale (hence $\dot{S}_1=0$). In the dissipative regime $\mathbf{J}$ is not constant of motion anymore, its magnitude and direction will change due to the gravitational radiation of the binary \citep{GerPLB2009}:
\begin{eqnarray}
    \dot{J}&=&\dot{L}(\mathbf{\hat{L}} \cdot \mathbf{\hat{J}})\\
    \mathbf{\dot{\hat{J}}}&=&\frac{\dot{L}}{J}(\mathbf{\hat{L}}-(\mathbf{\hat{L}}\cdot\mathbf{\hat{J}})\mathbf{\hat{J}}).
\end{eqnarray}

In the 2.5th PN order the direction of the dominant spin shows a slow secular change, with a quasi-periodicity on the top of it due to the spin-orbit precession. When we assume the jet is coupled to the spin of its emitter black hole, then the jet direction will also show a secular change with a periodicity on the top of it.

Due to Doppler boosting, an observer measures the apparent flux density $S_\mathrm{app}$ instead of the intrinsic flux density $S_\mathrm{int}$, such that
\begin{eqnarray}
S_\mathrm{app}(t)=S_\mathrm{int} \delta(t)^{n-\alpha},
\end{eqnarray}
where $n$ is a number between $2$ (straight jet) and $3$ (spherical jet), and $\alpha$ is the spectral index. The Doppler factor depends non-linearly on the inclination angle, therefore if the jet is turning towards us, first we do not see much change in the apparent flux density, then at small inclinations the apparent amplitude of the flux density variation due to the jet precession will be non-linearly amplified with time. Since we expect the $\gamma$-ray flux radiated by a single blob while the radio emission coming from a distribution of emitting regions in a jet with half-opening angle, we expect the time evolution of the amplification of the $\gamma$-ray flux and the radio flux density due to the relativistic beaming to be somewhat different. We will model multi-wavelength flux variations of J1048+7143 in the subsequent paper following the present one.

We note that assuming a more complex substructure of the major flares (e.g. the three-peaks model) would only slightly change our binary model, since we only need the center of the major flares to derive the post-Newtonian characteristics of the SMBBH.

\subsection{Gravitational Lifetime of the SMBH Binary}

With the first period specified, we use Eq.~(\ref{eq:determine_T_GW}) to determine the remaining time until the final coalescence of the binary supermassive black hole, provided that spin--orbit jet precession caused the (quasi-)periodic signals. 
Possible values of the merger time as function of the mass ratio of the binary are shown in Fig.~\ref{fig:mergerPredict}.
We investigate possible mass ratios between the typical mass ratio values of merging SMBBHs $q = 1/3$ and $q = 1/30$ \citep{GerPLB2009}, adopting a total mass of $10^{9.16 \pm 0.2} \, M_\odot$ \citep{Paliya2021} at redshift $z =1.15$ \citep{Polatidis1995}.
The predicted time of the binary merger (as observed from Earth) is depicted in green, with the uncertainty in the period causing the prediction to be a time range rather than an explicit date.  
The red area highlights invalid mass ratios ($q \gtrsim 1/4$) for the prediction due to the third major flare happening at that time. 
If a merger would have happened during the third flare, it would have either been long-lasting, as the jet precession period would have significantly changed, or stopped abruptly, as the merged black hole would have had a different jet direction than before the merger.
As it can be seen in Fig.~\ref{fig:mergerPredict}, the binary could merge in the next $\sim 60$ to $80$~yr, depending on the actual mass ratio.

With the remaining time until the merger determined, the previously used procedure with Eq.~(\ref{eq:determine_T_GW}) can be reversed: instead of calculating $\Delta T_{\rm GW}$ with a given period $P_{\rm jet} = P_{i\to(i+1)}$ between the flares $i$ and $(i+1)$, $\Delta T_{\rm GW}$ can be used to calculate the next period $P_{(i+1)\to(i+2)}$ between the detected flare $(i+1)$ and the next flare $(i+2)$. 
Since the remaining time until the final coalescence is a fixed value for the system and it only depends on the mass ratio, the next period can also be determined as a function of the mass ratio. 

For this calculation, we take advantage of the fact that the time until the final coalescence decreases by each newly determined period, as seen in Eq.~(\ref{eq:determine_T_GW}).
However, with each period, the influence of the half-opening angle $\zeta$ increases linearly, as the jet points at the observer every $360\degr \pm \zeta$.
 
We can illustrate this principle by predicting the period of an $n$-th flare with the known period of an $m$-th flare, with $m \in [1, n-1]$:
\begin{eqnarray}
    &&\phi\left(\Delta T_{\rm GW} - \sum_i^{m} P_{i\to(i+1)} \,,\, q\right) \pm m \zeta=\nonumber \\
    &&=\phi\left(\Delta T_{\rm GW} - \sum_i^{n} P_{i\to(i+1)} \,,\, q\right) \pm n \zeta \,.
    \label{eq:determine_Pn}
\end{eqnarray}

\subsection{Validation of the Model}
We first use the above model to confirm whether the time characteristics of the three major $\gamma$-ray flares of J1048+7143 are consistent with the scenario of a SMBBH merger at its core. For that, based on the first period determined, we predict the length of the second period and thus the occurrence time of the third flare in the framework of an inspiraling black hole binary.
This prediction is shown in Fig.~\ref{fig:flarePredict}.

In Fig.~\ref{fig:flarePredict}, the detected (gray) and predicted (blue) flares and their dependence on the binary mass ratio are presented. The width of the predicted time interval is always greater than the duration of the flare because we include the errors of the period and duration in our calculation, with the former increasing the range of valid next periods due to propagation of error.
The latter includes the half-opening angle of the jet $\zeta$, as it defines how long the jet can be detected. 
The prediction area thus confines the uncertainty band, inside which the next flare will occur.

Since the binary will merge later in time, if it has a small mass ratio (see Fig.~\ref{fig:mergerPredict}), the next precession periods decrease more slowly with time compared to the predicted periods at a larger mass ratio, as it then merges earlier. That is why the predicted time interval for the third flare is shifted to later times with a decreasing mass ratio. 

The uncertainty band for the predicted third flare (blue area with diagonal stripes going from bottom left to upper right) is completely in agreement with the actually detected one for most mass ratios, as can be seen with the overlapping gray and blue bands in Fig.~\ref{fig:flarePredict}. 
Only for mass ratios $q \gtrsim 1/5$ the prediction disagrees with the flare duration determined, disallowing such mass ratios for a binary at the center of J1048+7143 and extending the invalid area, marked in red, compared to Fig.~\ref{fig:mergerPredict}.
This overlap also means that the precession period predicted between the second flare detected and the predicted time of the third flare is consistent with the actual period determined, as this is the condition for the overlap.

The quasi-periodic major $\gamma$-ray flares from J1048+7143 are thus explainable with a precessing jet due to an ongoing merger of a SMBBH at its core. 
Because of the precession, the jet interacts periodically with a target field (matter or radiation), producing the observed $\gamma$-ray flares via proton-proton or proton-$\gamma$ interactions. 
As a consequence, a periodic neutrino emission is also expected in such a scenario.

\subsection{Prediction of the Next $\gamma$-ray Flare}

We use the successful prediction of the third $\gamma$-ray flare with our model in order to predict when the possible fourth $\gamma$-ray flare will occur.
For that, based on the procedure explained above, we take the first period determined as an input and evolve our model until the third potential period with Eq.~(\ref{eq:determine_Pn}). This means that the prediction of the third $\gamma$-ray flare with the second period is included in this calculation.

The uncertainty band for the fourth flare predicted with this method is shown in Fig.~\ref{fig:flarePredict} with the blue area with diagonal stripes going from upper left to bottom right. 
Its width is greater than that for the third flare prediction, as the error of one additional period is taken here into account due to propagation of error. 

As it can be seen, the uncertainty band for the fourth flare prediction  overlaps only partially with the third major flare detected at high mass ratios ($q \gtrsim 1/10$), since the subsequent periods at these mass ratios decrease significantly. However, since the fourth flare can occur sometime in the uncertainty band, mass ratios until the red band for invalid flare predictions of $q \lesssim 1/5$ are still a possibility. 

The vertical dashed line in Fig.~\ref{fig:flarePredict} marks the last day of available data from the $\gamma$-ray photon flux curve of J1048+7143 seen in Fig.~\ref{fig:gammalc}. 
No new flare was detected until this date, 2022 Mar 14, so that we can narrow down the beginning of the next (fourth) flare from J1048+7143. We note, the last two data points in Fig.~\ref{fig:gammalc} could indicate the beginning of a new flare.

The checkered violet area in Fig.~\ref{fig:flarePredict}. marks the projection of the third detected $\gamma$-ray flare (in gray) with the next period predicted onto the fourth flare prediction. 
This corresponds to the more probable prediction of the fourth flare, starting from the second detected flare, using the second period determined along with a fixed duration of the third detected flare (including its error bars). 
The benefit is that this way, the uncertainty band of the fourth flare prediction is narrowed down even further, as the error propagation of the period starting with the first flare is avoided. The violet uncertainty band in Fig.~\ref{fig:flarePredict} is still wider than the third flare duration including its errors because it includes the same duration error as well as an additional error in the third period prediction.

Based on the discussion above, we predict that the next, fourth flare from J1048+7143 will occur before the end of 2024 (violet band), at the latest before summer 2025 (blue band), if a currently ongoing SMBBH merger is indeed the cause of the quasi-periodic flares.
If the duration of it will overlap with the violet or blue prediction bands, then we can possibly narrow down the range of the valid mass ratios for the binary even further. 
\begin{figure*}
    \centering
    \includegraphics[angle=0,scale=0.5]{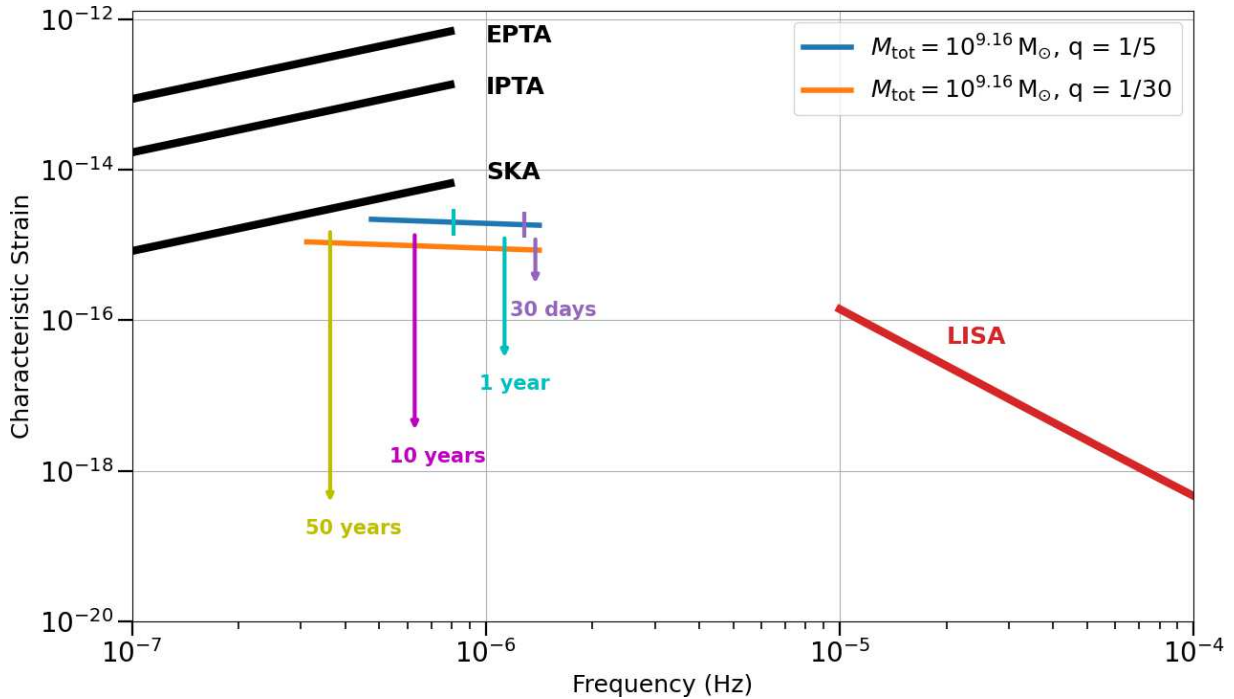}
    \caption{Gravitational Wave signals from J1048+7143 on the condition of a SMBBH merger at its core. Blue and orange lines: expected characteristic strain with the mass ratio $q=1/5$ and $q=1/30$ respectively.
    Black lines: sensitivity curves from the detectors SKA, IPTA and EPTA, constructed after \citet*{Moore2015}. Red line: LISA sensitivity curve, plotted following \citet*{Robson2019}. Vertical lines indicate the gravitational wave frequency emitted by a possible binary with (from left to right) 50 years (yellow), 10 years (magenta), 1 year (cyan), and 30 days (purple) until the merger for the respective mass ratio.}
    \label{fig:strain}
\end{figure*}
\section{Gravitational Wave Signal expected from J1048+7143}
\label{sec:gwrad}

\subsection{The Expected Characteristic Strain}
After having shown that the $\gamma$-ray signals from J1048+7143 are compatible with a SMBBH merger at its core, we can determine how its gravitational wave signal will look like. 
For that, we model the expected characteristic strain $h_c$ after \cite{Sesana2016}:
\begin{equation}
    h_c = \sqrt{\frac{2}{3}}
    \, \frac{1}{r(z)}
    \, \frac{1}{\pi^{2/3} c^{3/2} } 
    \, \frac{\left(G \mathcal{M}\right)^{5/6}}{(1 + z)^{1/2}}
    \, f^{-1/6} \,,
\end{equation}
with the the comoving source distance $r(z)$, the chirp mass $\mathcal{M} = (m_1 m_2)^{3/5}/(m_1 + m_2)^{1/5} $ and the observed GW frequency $f$.
The corresponding GW frequencies range from arbitrary small frequencies the binary can emit with its first gravitational radiation on the one side and $f_{\rm ISCO}$ on the other.
The latter describes the emitted GW frequency at the innermost stable circular orbit (ISCO) and it is defined as \citep{Vecchio2004}:
\begin{equation}
    f_{\rm ISCO} = 6^{-3 / 2} \cdot \frac{c^{3}}{\pi G M (1+z)} \,.
\end{equation}
The expected characteristic strain is plotted in Fig.~\ref{fig:strain} in blue for a mass ratio of $q=1/5$ and in orange with a mass ratio of $q=1/30$. 
Other mass ratios lie in between these limiting values of $q$.
We determine the frequencies at which gravitational waves are emitted from this source with remaining 50, 10 and 1 year as well as 30 days until the final coalescence after \cite{Vecchio2004} and \citet*{Berti2005}, and we indicate them with vertical dashes in the colors (from left to right) yellow, magenta, cyan, and purple in Fig.~\ref{fig:strain}. 
The leftmost frequencies, for which the characteristic strain is plotted, marks the current position (on 2022 Sep 9) of the binary in the diagram. It was determined with Fig.~\ref{fig:mergerPredict} by calculating the maximum time the binary needs to merge at the respective mass ratios ($\sim 6$ years for $q = 1/5$ and $\sim 77$ years for $q = 1/30$).
In the following, we test whether these signals could be detectable with current and planned pulsar timing arrays (PTAs) and LISA.

\subsection{Detectablility of the Expected GW Signal}

The current and future expected sensitivity curves from the European Pulsar Timing Array (EPTA), International Pulsar Timing Array (IPTA), and Square Kilometer Array (SKA) are shown in black \citep*{Moore2015} in Fig.~\ref{fig:strain}. They were constructed using 5(20) pulsars with 10(15) years of observation time for EPTA(IPTA), and 100 pulsars with 20 years of observation time for SKA. The LISA sensitivity curve is shown in red.

Figure~\ref{fig:strain} shows that neither the PTAs nor LISA can detect J1048+7143 in gravitational waves.
For IPTA and SKA, that is because the characteristic strain of the SMBBH at the leftmost frequencies (on 2022 Sep 9) lies for each mass ratio below their respective sensitivity curves.
Since the binary requires up to 77 years to merge with a mass ratio of $q=1/30$ and even less for a higher mass ratio (see Fig.~\ref{fig:mergerPredict}), it is currently in a stage where the emitted GW frequency is too large and the emitted characteristic strain is too low to be detected by the PTAs.

With the SKA, a SMBBH with the same estimated total mass and redshift as J1048+7143 could be detectable roughly up to $50$ years before merger for the mass ratio $q=1/5$ and roughly up to $500$ years before merger for the mass ratio $1/30$, with mass ratios in between lying in this time range, provided the signal is distinguishable from the background and noise.
However, as such a binary will not merge in the sensitivity range of SKA, the distinctive rise in signal frequency at the time of the merger will not be detected by SKA. Without it, a possible distinction of the strain from the stochastic gravitational wave background is at least challenging, in particular, since only a few periods of the gravitational signal can be detected at such frequencies. On top of that, possible detector noise could decrease the detector sensitivity at the relevant frequencies.

On the other side, LISA is aimed to detect SMBBH mergers with smaller masses than that of J1048+7143 ($\lesssim 10^8 \, M_\odot$, see \citealt{LISA2017_WP}) and is therefore not sensitive for this source because its expected GW frequency falls outside the frequency coverage of LISA.

\section{Summary and Conclusions}
\label{sec:sumconcl}

We analyzed Fermi-LAT observations of the flat-spectrum radio quasar 4FGL~J1048.4+7143 (J1048+7143). We performed the unbinned likelihood analysis of the data with adaptive binning in the $168\,\mathrm{MeV} - 800\,\mathrm{GeV}$ energy range, to generate the light curve of this FSRQ with a time coverage of $\sim13.6$ yr between 2008 Aug 4 and 2022 Mar 14.

We found that J1048+7143 has shown three major $\gamma$-ray flares in the studied time and energy range, such that all three flares can be divided into two sharp sub-flares. We applied the so-called centroid method on the exponential fit of the $\gamma$-ray light curve to define where the flares start and end. We found the elapsed time between major $\gamma$-ray flares one and two to be $P_{1\to2} = 3.19 \pm 0.24$~yr, and between the major $\gamma$-ray flares two and three to be $P_{2\to3} = 3.07 \pm 0.19$~yr. 

We investigated the arcsec-scale radio structure of J1048+7143 by analyzing archival VLA imaging data. We derived the structure of the VLBI jet by analyzing archival $8.6$-GHz VLBI calibrated visibility data taken in 69 epochs (between 1994.61 and 2020.73). We investigated the flux density variation of J1048+7143 by employing $4.8$-GHz data from the Nanshan (Ur) and the RATAN-600 radio telescopes (between 2014.04 and 2021.08), and at 15-GHz employing single dish public observations taken with the OVRO 40m Radio Telescope (between 2008.02 and 2011.99).

The VLA observations reveal the main kpc-scale jet pointing roughly towards west, extending to $\sim 15$~kpc projected distance.  VLBI observations show the sign of two jet channels at pc-scales, one pointing to east, and one pointing to south, extending to the maximum $\sim 17$~pc projected distance. The kpc and pc-scale jets are significantly misaligned with each other which is a revealing sign of jet precession. Two major radio flares of J1048+7143 are unfolded by the 4.8-GHz total flux density curve of the source. Comparing the single-dish radio flux density curves with the $E>168$~MeV $\gamma$-ray light curve in the Fermi-LAT time coverage (Fig.~\ref{fig:radio_flux_curves}), we found the radio flares to slightly lag behind the $\gamma$-flares.

Assuming the periodic $\gamma$-ray flares arise due to the spin--orbit precession of the jet-emitting SMBH and expanding the model for the precession of the jet introduced by \cite{deBruijn2020}, we determined the direction angle of the jet $\phi$ as a function of the remaining time until the binary merger $\Delta T_{\rm GW}$. We used this model to confirm that the three major flares detected are in agreement with a SMBBH merger at the core of J1048+7143. 

If the scenario of the currently ongoing SMBBH merger being consistent with the observations is indeed the cause of the quasi-periodic $\gamma$-ray flares, then the fourth $\gamma$-ray flare from J1048+7143 will be detected before the end of 2024, but before summer 2025 at the latest. Assuming the SMBBH scenario holds, the binary could merge in the next $\sim 60$ to $80$~yr, depending on the actual mass ratio.

With the $\gamma$-ray signals from J1048+7143 being compatible with a SMBBH merger at its core, we determined how its gravitational wave signal will look like. We found that neither the pulsar timing arrays EPTA, IPTA, nor the future SKA and LISA could detect J1048+7143 in gravitational waves.

In this paper, we presented the multimessenger picture of J1048+7143 which is consistent with $\gamma$-ray and radio observations, under the assumption that the observed light variation in the $\gamma$-ray and radio regimes, as well as the peculiar kpc- and pc-scale jet structure of J1048+7143 are due to the spin precession of the jet-emitting black hole at the heart of the host galaxy. Identification of sources similar to J1048+7143 has high importance to be able to reveal periodic neutrino factories in the distant Universe, from where we could observe only neutrinos and gravitational waves.

\begin{acknowledgments}
E.K. thanks the Hungarian Academy of Sciences for its Premium Postdoctoral Scholarship. J.T. acknowledges support from the German Science Foundation DFG, via the Collaborative Research Center \textit{SFB1491: Cosmic Interacting Matters - from Source to Signal}.
E.K., S.F., and K.\'E.G. were supported by the Hungarian National Research, Development and Innovation Office (NKFIH), grant number OTKA K134213.
L.C. was supported by the Chinese Academy of Sciences (CAS) ``Light of West China'' Program (No. 2021-XBQNXZ-005).
This paper makes use of publicly available Fermi-LAT data provided online by the \url{https://fermi.gsfc.nasa.gov/ssc/data/access/} Fermi Science Support Center. On behalf of Project 'fermi-agn' we thank for the usage of the ELKH Cloud that significantly helped us achieving the results published in this paper. 
The National Radio Astronomy Observatory is a facility of the National Science Foundation operated under the cooperative agreement by Associated Universities, Inc. 
We acknowledge the use of data from the Astrogeo Center database maintained by Leonid Petrov.
\end{acknowledgments}

\appendix
\section{Additional sources in the ROI about J1048+7143}
\begin{deluxetable*}{cccccccc}
\tablecaption{Additional sources in the ROI about J1048+7143.\label{table:addsources}}
\tablewidth{0pt}
\tablehead{
\colhead{Source ID} & \colhead{RA} & \colhead{DEC} & \colhead{$\Delta d$} & \colhead{$F_{1000}\times 10^{10}$} & \colhead{$\mathrm{err}F_{1000}\times 10^{10}$} & \colhead{TS} & \colhead{$N_\mathrm{pred}$}\\
\colhead{-} & \colhead{$(\degr)$} & \colhead{$(\degr)$} & \colhead{$(\degr)$} & \colhead{$(\mathrm{ph}~\mathrm{cm}^{-2}~\mathrm{s}^{-1}$)} & \colhead{$(\mathrm{ph}~\mathrm{cm}^{-2}~\mathrm{s}^{-1})$} & \colhead{-} & \colhead{-}}
\decimalcolnumbers
\startdata
PS J1055.5+6509 & 163.889 & 65.165 & 6.597 & 1.99 & 0.29 & 99.16 & 484.91\\
PS J1135.7+6612 & 173.942 & 66.211 & 6.939 & 1.06 & 0.24 & 36.33 & 320.16\\
PS J1207.8+6926 & 181.966 & 69.444 & 6.947 & 1.09 & 0.25  & 34.19 & 233.38\\
PS J1142.5+6412 & 175.635 & 64.216 & 9.021 & 1.28 & 0.28 & 56.46 & 434.91\\
PS J1244.1+7635 & 191.050 & 76.598 & 9.135 & 0.83 & 0.24 & 27.33 & 95.50\\
\enddata
\tablecomments{Source ID (1), right-ascension J2000 (2), declination J2000 (3), distance between the source and J1048+7143 (4),  flux measured by Fermi between $0.1$\,GeV and $100$\,GeV (5), error of the flux value (6), TS-value of the detection (7), predicted number of photons (8).}
\end{deluxetable*}

\end{document}